# Gas clumping in the outskirts of galaxy clusters, an assessment of the sensitivity of STAR-X

Christian T. Norseth[1]*, Daniel R. Wik,[1] John A. ZuHone,[2] Eric D. Miller,[3] Marshall W. Bautz[3] and Michael McDonald[4]

[1]*Department of Physics and Astronomy, University of Utah, 201 James Fletcher Building, 115 S. 1400 E., SLC, UT 84112-0830, USA*
[2]*Center for Astrophysics, Harvard and Smithsonian, 60 Garden St., Cambridge, MA 02138, USA*
[3]*Kavli Institute for Astrophysics and Space Research, Massachusetts Institute of Technology, 77 Massachusetts Avenue, Cambridge, MA 02139, USA*
[4]*Department of Physics, Massachusetts Institute of Technology, Cambridge, MA 02139, USA*

**ABSTRACT**
In the outskirts of galaxy clusters, entropy profiles measured from X-ray observations of the hot intracluster medium (ICM) drops off unexpectedly. One possible explanation for this effect is gas clumping, where pockets of cooler and denser structures within the ICM are present. Current observatories are unable to directly detect these hypothetical gas clumps. One of the science drivers of the proposed *STAR-X* observatory is to resolve these or similar structures. Its high spatial resolution, large effective area, and low instrumental background make *STAR-X* ideal for directly detecting and characterizing clumps and diffuse emission in cluster outskirts. The aim of this work is to simulate observations of clumping in clusters to determine how well *STAR-X* will be able to detect clumps, as well as what clumping properties reproduce observed entropy profiles. This is achieved by using yt, pyXSIM, SOXS, and other tools to inject ideally modelled clumps into 3D models derived from actual clusters using their observed profiles from other X-ray missions. Radial temperature and surface brightness profiles are then extracted from mock observations using concentric annuli. We find that in simulated observations for *STAR-X*, a parameter space of clump properties exists where gas clumps can be successfully identified using wavdetect and masked, and are able to recover the true cluster profiles. This demonstrates that *STAR-X* could be capable of detecting substructure in the outskirts of nearby clusters and that the properties of both the outskirts and the clumps will be revealed.

## 1 INTRODUCTION

Galaxy clusters are collections of 100s to 1000s of galaxies and are the largest gravitationally bound structures in the universe. The intracluster medium (ICM), mainly containing a hot ($\sim 10^7$ K) gas that permeates the space between galaxies in a cluster is detectable in X-rays via its thermal bremsstrahlung radiation, as well as radiation from collisional ionization. The thermodynamical properties of the ICM in galaxy clusters reflect the history of their growth and allow cluster masses to be estimated, which can be used to estimate cosmological parameters (Allen et al. 2011). It is therefore important that the properties derived from the thermodynamic state of the gas in galaxy clusters are accurate. Using X-ray telescopes, radial temperature, and surface brightness profiles of a cluster can be extracted from imaging and spectral analyses of an observation. The entropy of the gas, inferred from temperature and density estimates, is particularly essential to understanding the thermodynamical properties of a cluster, including its evolution (Ghirardini et al. 2021). However, observations have shown unexpected results in the derived entropy profiles for many galaxy clusters. The entropy $K$ is predicted to increase nearly proportional with radius ($K \propto r^{1.1}$) in a relaxed cluster in hydrostatic equilibrium (Voit et al. 2005). However, in the outskirts of galaxy clusters, entropy profiles measured from X-ray observations of the ICM drop off unexpectedly. This trend is seen in *Suzaku* observations of cluster outskirts, such as in the Perseus Cluster (Simionescu et al. 2011), Abell 1246 (Sato et al. 2014), Abell 2029 (Walker et al. 2012a), Abell 1795 (Bautz et al. 2009), Abell 1835 (Ichikawa et al. 2013), and many others (Walker et al. 2019).

One possible explanation for this unexpected drop in entropy is gas clumping, where pockets of cooler and denser structures within the ICM are present in the outskirts. Entropy is typically estimated from deprojected temperature $T(r)$ and density $n(r)$ profiles, based on fits to projected temperature and surface brightness profiles, as

$$K(r) = \frac{T(r)}{n(r)^{2/3}} . \qquad (1)$$

Pockets or clumps of cooler and denser gas more apparent at larger radii would be capable of biasing the projected temperature and especially surface brightness ($\propto n^2$) measurements to lower and higher values, respectively, driving the inferred $K(r)$ below the true entropy of the ambient ICM. Since the properties of gas clumps are unknown, it is therefore a combination of temperature and density bias that drives the entropy down, whether it be mostly through a bias in the density profile or a significant bias in the temperature profile as well. While the low and stable background of *Suzaku* allowed thermodynamic profiles to be measured at large radii, *Suzaku*'s spatial resolution was insufficient to allow the clumps, if they exist, to be resolved and directly detected. This concept of gas clumping is supported by simulations, which have shown that the distribution of the density of the ICM in the outskirts tends to be non-uniform (Nagai & Lau 2011). While interfering with the ability to accurately measure entropy, gas clumping in the outskirts is also of interest due to its likely role in the physics and evolution of galaxy clusters (Walker & Lau 2022). It is therefore important for gas clumps to be directly identified, characterized, and ultimately masked out to recover the true entropy profile of the diffuse gas and properly test our current understanding of the evolution of the ICM in galaxy clusters.

*E-mail: u1033962@utah.edu

Currently, gas clumps in the outskirts of clusters have not been clearly detected by sufficiently sensitive observatories, such as the XIS (X-ray Imaging Spectrometer) instrument on the *Suzaku* telescope. The lower background provided by Suzaku's low-Earth orbit allowed it to explore the diffuse emission in cluster outskirts, which neither *Chandra*, *XMM-Newton*, nor *eROSITA* can take advantage of. The superior spatial resolution of the *Chandra* and *XMM-Newton* observatories hold more promise in principle, but the higher and less stable background rates have complicated probes of cluster outskirts. It is also unlikely that *eROSITA* would detect gas clumps as its on-axis PSF is comparable to XMM Newton's.

There have been methods developed that correct for bias due to clumping in derived emissivity profiles (Eckert et al. 2015; Eckert et al. 2017). Using the azimuthal median of concentric annuli in the surface brightness of cluster emission, the expected entropy profile can be recovered to within $1\sigma$ (Eckert et al. 2015). Although clumps have not been directly detected, it is clear that correcting for inhomogeneities in the ICM is necessary to properly extract the density and therefore entropy profile of a cluster (Morandi & Cui 2013).

Although we can roughly account for gas clumping, directly detecting clumps would allow for a much more accurate derivation of cosmological parameters, as well as reinforcing our understanding of cluster evolution. In addition, the median method mentioned above is limited by spatial resolution.

A potential solution to this limitation is the *Survey and Time-domain Astronomical Research eXplorer* (*STAR-X*). *STAR-X* was proposed to the Mid-Sized Explorer Class Mission (MIDEX) call by NASA at the end of 2021 and was selected for Phase A study. Like *Suzaku*, *STAR-X* will have a low instrumental background due to its low earth orbit, where it will be protected from high energy charged particles by the Earth's magnetosphere. This low background at harder ($E > 2$ keV) energies, coupled with a high soft effective area, will allow sensitive measurements of the outskirts of galaxy clusters. The design of *STAR-X* provides a large (1° diameter) field of view (FOV), large effective area, and a 2 arcsec half-power diameter point-spread function (PSF) to minimize confusion with point sources and efficiently survey the outskirts of nearby clusters in single, ∼100 ks observations (Zhang 2017). While the high, soft ($E < 2$ keV) effective area allows cooler gas clumps to be directly imaged and masked, the low particle background at higher energies permits accurate temperature measurements of the truly diffuse gas. These advances, possible with *STAR-X*, will provide a greater understanding of cluster evolution and clusters' connection to the cosmic web. In the following, we simulate *STAR-X* observations of a handful of nearby galaxy clusters with measured drops or flattening in their entropy profiles. We assume the true diffuse gas follows the expected entropy relation and inject clumps with various properties that would have gone undetected by *Suzaku* and reproduce the observed entropy profiles. The mock observational data are then realistically analysed to determine if *STAR-X* would be able to detect gas clumping in galaxy cluster outskirts and recover the injected diffuse entropy profile.

Throughout this work, we assume a flat $\Lambda$CDM cosmology with $H_0 = 71$ km s$^{-1}$ Mpc$^{-1}$, $\Omega_m = 0.27$, $\Omega_\Lambda = 0.73$. Unless otherwise stated, uncertainty ranges are reported as 90 per cent confidence intervals.

## 2 METHODS

### 2.1 Simulating *STAR-X* observations

In order to assess the impact of gas clumping and the sensitivity of *STAR-X*, a simulated observation or event list of a reconstructed galaxy cluster must first be produced.

A summary of the simulation steps are given below:

(i) Temperature and emissivity profiles for a given cluster are defined, based on literature fits to observations of that cluster.

(ii) The uniformly sized cells of a data cube are populated with values based on the profiles.

(iii) Gas clumps are randomly injected into the data cube based on their assumed properties.

(iv) A 3D photon list that represents the emission of each cell is produced.

(v) The photon list is projected into a 2D event file.

(vi) Point sources consistent with the cosmic X-ray background (CXB) are added with random positions to the event file.

(vii) Photons representing the Galactic foreground and *STAR-X*'s estimated instrumental background are also added with random positions, presuming a uniform spatial distribution of both components.

We model the radial temperature and emissivity profiles with the following form, as presented in Vikhlinin et al. (2006):

$$T_{3D}(r) = T_0 \frac{\left(\frac{r}{r_{\text{cool}}}\right)^{a_{\text{cool}}} + \frac{T_{\text{min}}}{T_0}}{\left(\frac{r}{r_{\text{cool}}}\right)^{a_{\text{cool}}} + 1} \frac{\left(\frac{r}{r_t}\right)^{-a}}{\left[1 + \left(\frac{r}{r_t}\right)^b\right]^{c/b}}; \quad (2)$$

$$n_p n_e(r) = n_0^2 \frac{\left(\frac{r}{r_c}\right)^{-\alpha}}{\left(1 + \frac{r^2}{r_c^2}\right)^{3\beta - \alpha/2}} \frac{1}{\left[1 + \left(\frac{r}{r_s}\right)^\gamma\right]^{\epsilon/\gamma}}$$

$$+ \frac{n_{02}^2}{\left(1 + \left(\frac{r}{r_{c2}}\right)^2\right)^{3\beta_2}}. \quad (3)$$

The radial distance $r$ determines how the temperature $T_{3D}$ and emission measure ($n_p n_e$) varies inside the model cluster, assuming spherical symmetry, controlled by radial scale parameters $r_{\text{cool}}$, $r_t$, $r_c$, $r_s$, and $r_{c2}$ and dimensionless parameters $a_{\text{cool}}$, $a$, $b$, $c$, $\alpha$, $\beta$, $\epsilon$, $\gamma$, and $\beta_2$. In equation (2), the profile is normalized by $T_0$ and modulated inside the cool core by $T_{\text{min}}$, while the emissivity is normalized by $n_{02}$ in the centre and $n_0$ elsewhere in equation (3). The terms controlling the profiles in the outskirts are adjusted so that the entropy profile roughly follows the $r^{1.1}$ relation; clumps are later added to match the observed entropy profile. See Section 2.3 for more details on the entropy correction. The parameter values for the emission measure and temperature functions are given in Tables 1 and 2, respectively. In order to produce mock observations, we first create discrete 3D grids with cell size based on a grid $512^3$ cells in total size that extends out to the virial radius of each cluster. Our three clusters, Abell 2029 (see Section 4.1 for more details), Abell 1246 (Section 4.2), and the Perseus cluster (Section 4.3) were recreated out to ∼2.3 Mpc in radius, corresponding to cell sizes of 8.9, 8.6, and 8.6 kpc, respectively. These data cubes are then populated with temperature and emissivity values from equations (2) and (3) to represent the truly diffuse component of the ICM.

In Step 3, clumps are injected into the data grids. The number of clumps, their central temperature, central density, and radius are chosen from a predetermined matrix of values (see Section 3.1 for more details). Each clump is built from the centre outwards to create a spherical area in the data grid with a temperature and density profile $\rho(r)$ with a Gaussian form defined by the radius $R_{cl}$ encompassing its full width at half-maximum (FWHM), such that

$$\rho(r) = \rho_0 e^{-r^2/2\sigma^2}, \qquad (4)$$

where the standard deviation $\sigma = R_{cl}/\sqrt{2\ln 2}$. The clumps are identical and randomly distributed in the 3D space of the grids. While the true clump distribution is unknown and may only exist in cluster outskirts, clumps that fall within $R_{500}$ will contribute negligibly to the emission there; a uniform placement is thus the simplest, realistic distribution to choose. Clump properties for a range of assumed sizes $R_{cl}$ that reproduce observed entropy profiles are given in Table 3.

The temperature and emissivity data cubes are then loaded into yt.[1] In combination with the cluster's radius, which is used to generate a cubic bounding box with dimensions equal to the diameter of the cluster near its virial radius, a yt data set can be generated that stores the temperature and density values within the cells of a 3D data set. The data set, combined with the redshift, position, collecting area, and exposure time for the mock observation, is used to generate a 3D photon list based on an apec thermal source model using the python package pyXSIM.[2] The 3D photon list is then projected along a chosen axis to produce a 2D event file. An absorption model (tbabs) is also included in this step to account for the hydrogen column $N_H$ in the direction of the cluster.

Lastly, the 2D event file is provided as input to SOXS,[3] an instrument simulator. SOXS simulates the response and other properties of *STAR-X*, including the FOV, the chip and pixel sizes and arrangements, the focal length, the PSF, the effective area, and the instrument's particle background. Point sources are also added into the simulated observation by SOXS, based on the log *N*–log *S* flux distribution of CXB sources measured in Lehmer et al. (2012) from the *Chandra* Deep Field South observations. Using SOXS, a mock observation of the reconstructed cluster is finally produced and can be analysed using standard tools and techniques.

### 2.2 Mock observation analysis

At this stage, the mock event file is analysed using existing tools designed for observatories that obtain event lists with spectro-imaging information, such as *Chandra*. Knowledge of the locations or properties of CXB sources (except where noted below) and clumps are not used at any point in the analysis. The analysis steps are as follows.

(i) Point sources are identified using wavdetect,[4] a Chandra Interactive Analysis of Observations (CIAO)[5] tool, and are masked to produce a point-source-corrected image. A separate point-source-corrected image is produced using known point source positions attained from a simulated image that does not include clumping, but does include the same, randomly generated point sources. This second point-source-corrected image is used to assess the accuracy of the initial point-source-correction where clumps are included. wavdetect uses 'Mexican Hat' wavelet functions with different scale sizes to identify sources. A smaller scale size equivalent to the approximate size of a point source is provided to wavdetect in this step.

(ii) Clumps are then identified using wavdetect and subtracted out to produce a point source and clump corrected image. A scale size equivalent to the approximate size of a gas clump is provided to wavdetect in this step.

(iii) Spectra are then extracted from concentric annuli centred at the X-ray emission peak, or the centre of the cluster. Annuli are chosen to have roughly the same number of counts per region. Background spectra are also extracted from a simulated, blank sky image. The blank sky image includes galactic foreground, instrumental background, and point sources. The point sources are simulated in the same positions as the cluster observation and masked out using the point source regions previously identified.

(iv) The spectra are fitted in XSpec using a standard apec * tbabs model to extract projected temperature values from each annulus. A radial temperature profile for the cluster is then produced.

(v) Using pyproffit,[6] surface brightness measurements are carried out using images produced from the 2D event files. pyproffit generates a deprojected, radial density profile $n(r)$ using the standard Onion-Peeling method. The annuli used in the spectral extraction are broken into smaller pieces in this step to allow for a greater resolution in the deprojected density data. Each annulus holds roughly the same amount of counts.

(vi) $T_{3D}$ is fit to the projected temperature profile and a radial entropy profile can then be derived using the equation

$$K(r) = \frac{T_{3D}(r)}{n(r)^{2/3}}. \qquad (5)$$

### 2.3 Entropy correction

In order to evaluate the impact of individual gas clumping properties, a cluster recreation must first be entropy-corrected, with the goal of injecting gas clumps in to drive the entropy profile down to match observations. Throughout this work, two central temperatures for gas clumps were evaluated, 3.0 and 0.7 keV. The former (3.0 keV) was chosen as it has a lesser impact on a cluster's temperature profile since the temperature in the outskirts for the chosen clusters is around 3.0 keV. This allowed for an analysis of a warmer, central clump temperature and only required a modification to the cluster's density profile to correct for entropy.

For clumps with a central temperature of 3.0 keV, the cluster's density profile was lowered in the outskirts to roughly line up with the normalized, expected entropy relation ($r^{1.1}$) found in Voit et al. (2005).

---

[1] https://yt-project.org
[2] http://hea-www.cfa.harvard.edu/jzuhone/pyxsim/index.html
[3] https://hea-www.cfa.harvard.edu/soxs/index.html
[4] https://cxc.cfa.harvard.edu/ciao/ahelp/wavdetect.html
[5] https://cxc.cfa.harvard.edu/ciao/
[6] https://pyproffit.readthedocs.io/en/latest/#

**Table 1.** Density profile reconstruction parameters. Modified profiles that correct for entropy are listed below the corresponding cluster. Profiles intended for gas clumps with a central temperature of 3.0 keV have only a modified density profile. For 0.7 keV clumps, the complementary, modified temperature profile can be found in Table 2.

| Cluster | $R$ (kpc) | $n_0$ ($10^{-3}$ cm$^{-3}$) | $r_c$ (kpc) | $r_s$ (kpc) | $\alpha$ | $\beta$ | $\epsilon$ | $n_{02}$ ($10^{-1}$ cm$^{-3}$) | $r_{c2}$ | $\beta_2$ |
|---|---|---|---|---|---|---|---|---|---|---|
| A2029[1] | 2300 | 15.721 | 84.2 | 908.9 | 1.164 | 0.545 | 1.669 | 3.510 | 5.00 | 1.0 |
| Entropy correction (3.0 keV) | ... | ... | ... | 1309 | ... | ... | 6.907 | ... | ... | ... |
| Entropy correction (0.7 keV) | ... | ... | 73.2 | 1309 | ... | 0.5 | 6.907 | ... | ... | ... |
| A1246[2] | 2200 | 0.300 | 5.0 | 2153 | 100.1 | 0.45 | 5.0 | 0.2 | 1.54 | 1.0 |
| Entropy correction (3.0 keV) | ... | ... | ... | 1953 | ... | ... | 17 | ... | ... | ... |
| Entropy correction (0.7 keV) | ... | ... | ... | 1910 | ... | 0.445 | 10 | ... | ... | ... |
| Perseus[3] | 2200 | 3.170 | 150 | 370.0 | 2.8 | 0.18 | 3.1 | 0.762 | 33.84 | 5.0 |
| Entropy correction (3.0 keV) | ... | ... | ... | 970.0 | ... | 0.3 | 7.1 | ... | ... | ... |
| Entropy correction (0.7 keV) | ... | ... | ... | 970.0 | ... | 0.348 | 5.0 | ... | ... | ... |

*Notes.* [1] Taken directly from Vikhlinin et al. (2006).
[2] Adapted from Sato et al. (2014).
[3] Adapted from Urban et al. (2013).

**Table 2.** Temperature profile reconstruction parameters. Modified profiles that correct for entropy are listed below the corresponding cluster, which are intended for gas clumps with a central temperature of 0.7 keV. Their complementary, modified density profiles can be found in Table 1.

| Cluster | $T_0$ (keV) | $r_t$ (Mpc) | a | b | c | $T_{min}/T_0$ | $r_{cool}$ (kpc) | $n_{02}$ $a_{cool}$ |
|---|---|---|---|---|---|---|---|---|
| A2029[1] | 16.19 | 3.04 | −0.3 | 1.57 | 5.9 | 0.10 | 93 | 0.48 |
| Entropy correction (0.7 keV) | 15.19 | ... | ... | 1.7 | 5.1 | ... | ... | ... |
| A1246[2] | 8.0 | 3.1 | 0.05 | 2.3 | 10.0 | 0.21 | 4.5 | 0.2 |
| Entropy correction (0.7 kev) | 8.5 | 3.9 | ... | ... | 9.0 | ... | 5.5 | 0.15 |
| Perseus[3] | 5.3 | 1.1 | 0.18 | 5.71 | 1.29 | 0.88 | 201 | 6.98 |
| Entropy correction (0.7 keV) | 5.6 | ... | 0.14 | 7.71 | 0.9 | ... | ... | ... |

*Notes.* [1] Taken directly from Vikhlinin et al. (2006).
[2] Adapted from Sato et al. (2014).
[3] Adapted from Urban et al. (2013).

**Table 3.** Detectable limits for gas clumps with central temperatures of 0.7 and 3.0 keV based on angular size and density in relation to *STAR-X*'s instrumental background and the galactic foreground. Densities listed are for clump photons that are at least 95 per cent detectable in an average of 10 samples. Number density refers to the number density of clumps within each cluster that are required to reproduce observed drops in entropy.

| | | Abell 2029 | | Abell 1246 | | Perseus cluster | |
|---|---|---|---|---|---|---|---|
| Radius $R_{cl}$ (kpc) | T (keV) | $\rho_0$ ($10^{-27}$ g cm$^{-3}$) | Number density ($10^{-8}$ kpc$^{-3}$) | $\rho_0$ ($10^{-27}$ g cm$^{-3}$) | Number density ($10^{-8}$ kpc$^{-3}$) | $\rho_0$ ($10^{-27}$ g cm$^{-3}$) | Number density ($10^{-8}$ kpc$^{-3}$) |
| 5 | 3.0 | 14.5 | 6.67 | 35.8 | 2.18 | 5.20 | 97.0 |
|   | 0.7 | 17.0 | 5.69 | 46.1 | 1.48 | 6.7 | 65.9 |
| 10 | 3.0 | 7.70 | 3.52 | 15.8 | 1.34 | 2.81 | 37.0 |
|    | 0.7 | 9.2 | 2.94 | 19.7 | 0.942 | 3.6 | 25.3 |
| 15 | 3.0 | 4.60 | 2.94 | 8.95 | 1.30 | 1.75 | 21.5 |
|    | 0.7 | 5.75 | 2.35 | 11.8 | 0.863 | 2.2 | 15.0 |
| 20 | 3.0 | 3.30 | 1.90 | 6.27 | 1.07 | 1.30 | 11.4 |
|    | 0.7 | 4.0 | 1.57 | 8.2 | 0.716 | 1.63 | 7.94 |
| 30 | 3.0 | 2.05 | 1.61 | 4.00 | 0.667 | 0.90 | 5.38 |
|    | 0.7 | 2.8 | 1.18 | 5.39 | 0.432 | 1.5 | 2.83 |
| 40 | 3.0 | 1.60 | 1.10 | 3.05 | 0.507 | 0.70 | 4.44 |
|    | 0.7 | 2.0 | 0.883 | 3.81 | 0.353 | 1.24 | 2.19 |
| 50 | 3.0 | 1.54 | 0.785 | 2.30 | 0.391 | 0.46 | 2.98 |
|    | 0.7 | 1.54 | 0.785 | 3.2 | 0.245 | 0.80 | 1.49 |

The overall normalization is adjusted to match the published data for each cluster in the radial range 0–2300 kpc, which results in factors of 1, 0.8, and 0.7 for Abell 2029, Abell 1246, and the Perseus Cluster, respectively.

For 0.7 keV central temperature clumps, both the temperature and density profiles of the cluster needed to be modified to correct for entropy as the cooler clumps bias the temperature profile down. This required a more complex correction, which relied on the results for gas clumping with a central temperature of 3.0 keV.

To start, a clump size was chosen from the detectable density analysis (Section 3) for gas clumps with a central temperature of 0.7 keV. Using the results from the entropy-corrected analysis of clumps with a central temperature of 3.0 keV (see Section 4 for more details), a total mass estimate for injected clumps that bias the cluster's density profile to match observations was calculated. This mass estimate relied on the entropy-corrected density profile used in the 3.0 keV clump analysis, where only the density profile was modified to correct for entropy. Using this mass estimate as a starting point, a number density of clumps with a central temperature of 0.7 keV for a chosen central clump radius could be calculated using the previously determined central clump density in the 0.7 keV detectability analysis (Section 3). With the chosen clump size, predetermined detectable density, and calculated number density, clumps were injected into a cluster recreation to observe the impact on the cluster's temperature profile. The simulated observations were then analysed to extract temperature and density profiles. A modified temperature profile could then be constructed to account for the predicted temperature drop. To achieve this, the temperature in each radial bin was increased by the difference between the cluster profile fit to observations and the extracted temperature profile with 0.7 keV clumps injected. The modified temperature profile was then used to re-simulate the cluster and adjusted until the injected clumps lowered the temperature profile down to match observations.

With the modified temperature profile, the original corrected density profile required modification to properly correct for entropy. The modified density profile was adjusted to account for the now raised temperature so that when combined, the expected entropy was approximated.

Clumps were then injected into the cluster with both profiles modified. The profiles and the number density of clumps were then adjusted to identify a combination of temperature and density parameters and a number density that resulted in: a combination of modified temperature and density profiles that resulted in the expected entropy, that when the clumps were injected they biased the profiles to match observations, and because the clump central density came from the detectability analysis, the clumps can be detected, masked out, and the modified profiles and expected entropy can be recovered. Thus, a possible combination of an entropy-corrected temperature and matching density profile was produced that corresponds with the detectable clump properties identified in Section 3.

### 2.4 Example: the *STAR-X* simulation of Abell 2029

The galaxy cluster Abell 2029 was initially selected for reconstruction as it had been observed to have a significant drop in entropy in its outskirts (near the virial radius), and the entire cluster fits within *STAR-X*'s 1° FOV. The 3D reconstruction of the cluster used temperature and emissivity profiles that were modelled using best fit parameters taken from Vikhlinin et al. (2006), based on the analysis of *Chandra* observations. Mock observations for *STAR-X*, as well as an approximation of the XIS Instrument on the *Suzaku* observatory, were generated and analysed to evaluate the accuracy of the simulation and analysis process. The projected temperature, deprojected density, and calculated entropy profiles derived from a recreation of Abell 2029 are presented in Fig. 1. The data produced in the analysis of the simulated images for both *Suzaku* and *STAR-X* follow the temperature and emissivity profiles that were used to construct the cluster, as well as observational data. This establishes that the simulation process and subsequent analysis process is able to reproduce observational data. With a reliable simulation and analysis process established, the clusters' profiles can now be adjusted to correct for entropy and cluster recreations can be injected with gas clumps to observe their impact and assess their detectability. The clump detectability analysis is described in detail in Section 3 and the individual cluster analysis can be found in Section 3.2.

An example of an entropy-corrected, *STAR-X* recreation of Abell 2029 with clumps injected, wavdetect regions overlaid, a point-source-masked and background-corrected image, and a spectrum extracted from an annulus of Abell 2029 is shown in Fig. 2. The corresponding profiles can be found in Fig. 4.

## 3 CLUMP PROPERTIES

Since the nature of clumps, if they even exist, is unknown, the range of properties they can have is fairly broad, with the main constraint being that they bias entropy profiles in cluster outskirts. Presuming a size range of individual clumps explored in simulations, we first determine how faint such clumps can be and still be separated from truly diffuse emission. The number density of clumps in the cluster volume is then set by the total amount of emission needed to produce drops in entropy comparable to those seen in actual observations. Individual clumps that are brighter (and fewer in total number) will be trivially detected by *STAR-X*, while fainter clumps would need to be more numerous to bias the entropy by the amount observed.

### 3.1 Properties of detectable clumps

Two uniform central clump temperatures were assessed; 3.0 and 0.7 keV.

A central temperature of 3.0 keV was used in order to limit the number of variables involved when finding number density values for clumps required to bias entropy profiles. This temperature of 3.0 keV for simulated gas clumps can be found in Vazza et al. (2012) with a high differential distribution and does not impact the projected temperatures of the cluster greatly as the outskirts are around 3.0 keV. This results in an entropy profile greatly impacted by the density of the clumps.

A central temperature of 0.7 keV was also assessed to examine cooler clumping scenarios, where the temperature profile of the cluster is also effected.

Before injecting clumps into cluster recreations, a range of clump sizes in kpc were selected from Vazza et al. (2012), chosen to match clump sizes found in cosmological simulations. Each clump size is evaluated separately, where individual cluster simulations are populated with clumps of identical size. For each radial size $R_{cl}$, a lower limit for detectable density was identified based on *STAR-X*'s instrumental background and the galactic foreground, at the redshift of each recreated cluster. The purpose of identifying these detectability limits is to test the limit of *STAR-X*'s resolving power. These detection limits allowed for an investigation into clumping properties that recreate observed drops in entropy, assuming that clumps are as faint as possible while still being at least 95 per cent detectable by *STAR-X*. The ICM is not included in this detectability analysis in order to evaluate the clumps against a constant background. When they are injected into their respective clusters, the clumps can be identified and subtracted out to fully recover the unbiased entropy profile, which indicates the clumps are fully detectable and it is appropriate to say they are along this arbitrary detectability threshold.

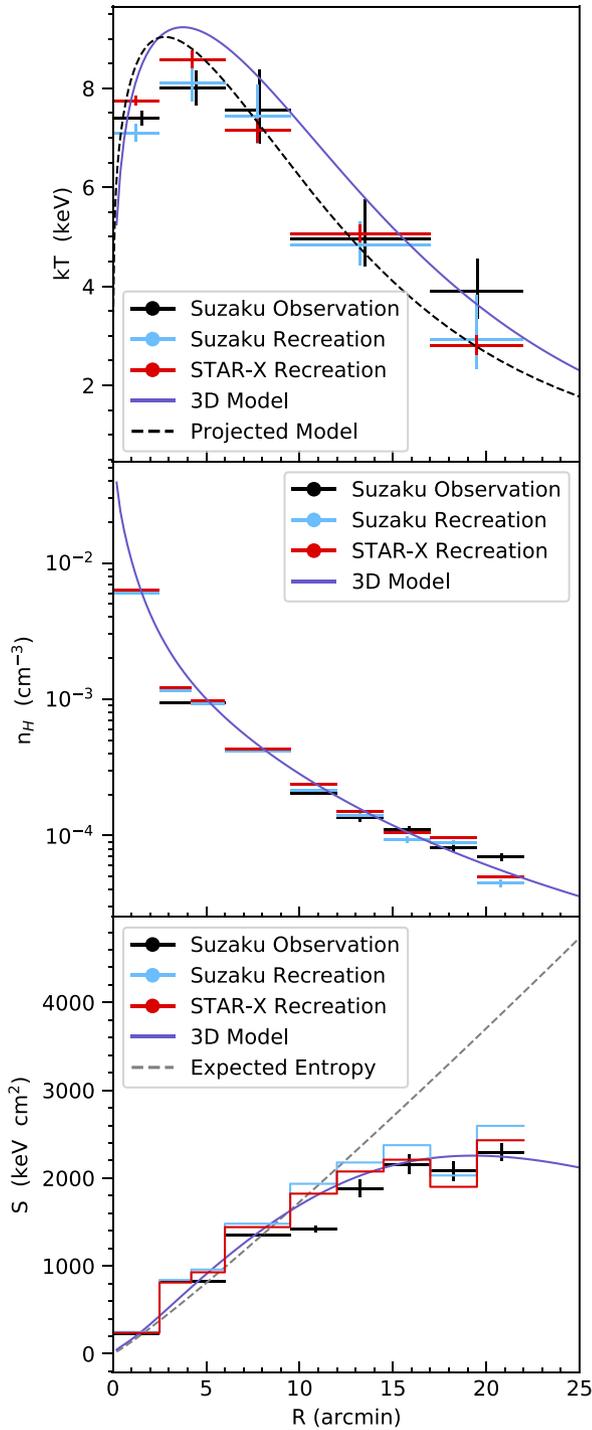

**Figure 1.** Top: Projected temperatures for Abell 2029 recreations compared to deprojected temperature values from *Suzaku* observations (Walker et al. 2012a). The dashed black line signifies the projected temperature profile. Middle: Deprojected hydrogen number density profiles. Bottom: Entropy profiles, dashed grey line signifies the expected entropy ($r^{1.1}$) relation. *Suzaku* observational data are shown in black, a simulated recreation of a *Suzaku* observation is shown in blue, and a simulated observation from *STAR-X* is seen in red. The solid purple lines signify best-fitting models used to reconstruct cluster.

The process for identifying detectability thresholds based on density and radial size is fairly straightforward. Gas clumps are randomly injected into an empty, spherical volume with a radius equal to the viral radius of a chosen cluster. The size of the gas clumps for each trial are assumed to be the same, with sizes spanning the range $5\,\mathrm{kpc} \leq R_{\mathrm{cl}} \leq 50\,\mathrm{kpc}$. Their corresponding angular size is determined by the redshift of each cluster.

Using yt, pyXSIM, and SOXS, photons corresponding to each clump in the volume are generated. The photons are then projected, and a mock observation of the clumps as seen by *STAR-X* are produced. Both an observation that includes *STAR-X*'s instrumental background and the Galactic foreground as well as one that is free of these backgrounds is produced. The observations are then analysed using wavdetect, where the approximate radius of the clump in pixel size is provided to wavdetect, calculated by converting the physical size of the clump to pixel coordinates given the redshift of the cluster. Clump-masked images are then produced and compared to determine the percentage of clump photons that were excluded in the observation with backgrounds relative to the observations without backgrounds included. The central gas density of the clumps is lowered incrementally to identify the threshold at which 95 per cent of the clump photons are detected. The density that satisfies this threshold is established as the baseline of detectability for the 'faintest' clumps.

The same procedure is carried out to identify clumps where only 5 per cent of clump photons are detected in order to establish the properties clumps need to have to avoid detection by *STAR-X*.

It is in this fashion that a lower limit for detectable clump density can be established over a range of constant, physical clump sizes for different clusters. The results from this initial density threshold analysis for Abell 2029, Abell 1246, and the Perseus Cluster can be seen in Table 3 and Fig. 3.

With a baseline for faintest detectable clump densities over a range of clump sizes established, the number density of clumps that recreate observed drops in entropy can now be investigated.

### 3.2 Recreating observed drops in entropy with clumps

With detection limits identified for a range of clump sizes, clumps can be injected into 3D reconstructions of observed clusters. These simulations are then used to determine the number density of clumps at the detectability threshold needed to replicate the observed drops in entropy for each cluster. The clumps that are injected into the reconstructions use the central density detection threshold values for each clump size as reported in Table 3 for a central clump temperature of 3.0 and 0.7 keV. All clumps injected into each reconstruction are identical in size, temperature, and central density. As described in Section 2.3, the observed density profile for the chosen cluster is adjusted in the outskirts to yield the expected entropy profile ($\propto r^{1.1}$) without clumping. Clumps are then injected into the 3D recreation and a mock observation is produced and analysed. For each clump size $R_{\mathrm{cl}}$, an increasing number of clumps is randomly injected into the volume until the extracted density and entropy profiles (before clump detection and removal) reflect those of real observations. In this way, the number of clumps needed to reproduce observed entropy drops in these clusters is determined.

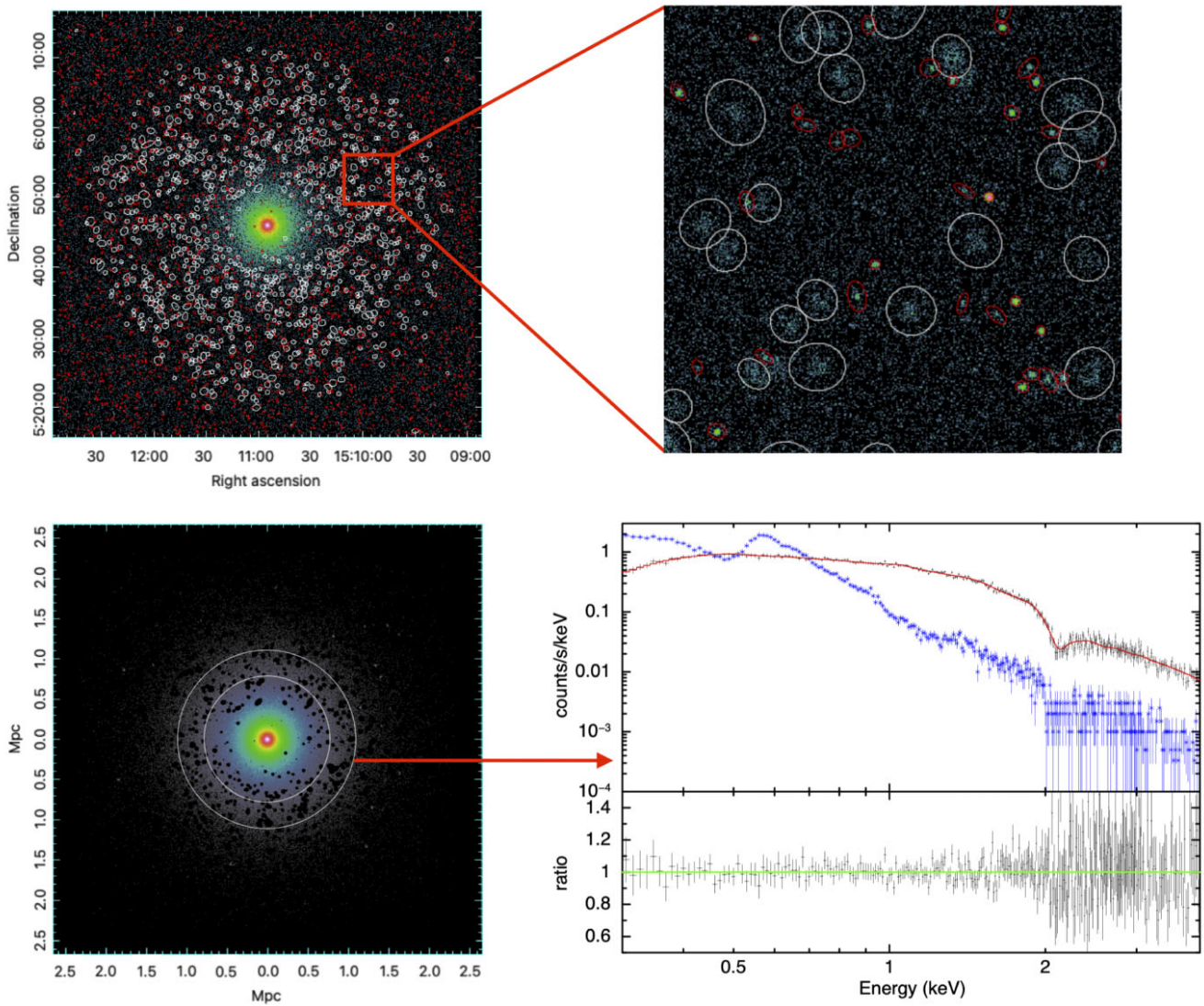

**Figure 2.** Top left: Simulated 1° × 1° observation of Abell 2029 with clumps injected, identified point sources circled in red and identified clumps circled in white. Top right: Zoomed-in region indicated by the red box in the image on left. Bottom left: Smoothed, background-subtracted version of the image above with point sources and clumps masked out. Both the top and bottom left cluster images are scaled logarithmically and have the same FOV. Bottom right: Fit of the spectrum extracted from Annulus 6, shown in white in the bottom left panel. The background-subtracted data are shown in black with the red curve representing the best-fitting model. The blue points indicate the background level. The lower panel shows the ratio of the data to the model, indicating the quality of the fit.

## 4 INDIVIDUAL CLUSTER SIMULATIONS

### 4.1 Abell 2029

As mentioned in section 2.4, the emissivity and temperature profiles used to recreate Abell 2029 were taken from Vikhlinin et al. (2006). See Tables 1 and 2 for the chosen density and temperature parameters initially used to reconstruct the cluster. Two central clump temperatures were assessed; 3.0 and 0.7 keV. For the following recreations, the temperature and density profiles were modified to correct for entropy. For gas clumps with a central temperature of 3.0 keV, only the density profile was lowered in the outskirts to yield the expected entropy profile. For gas clumps with a central temperature of 0.7 keV, both the temperature and density profiles were modified. For more details on the entropy-correction technique, see Section 2.3. The entropy-corrected profile parameters for both clump temperatures can be found in Tables 1 and 2.

For each central clump temperature (3.0 and 0.7 keV), each identified radial size and lower density limit established in Section 3.1 were injected into the cluster and a mock 100 ks *STAR-X* observation of the cluster was produced and analysed to extract a temperature and density profile following the methods outlined in Section 2 and compared to observations. For gas clumps with a central temperature of 3.0 keV, the number of clumps were increased until the density profile matched observations and the observed drop in entropy was recreated. Thus, a clump number density that recreated observations for each clump size and central gas density were identified, which are listed in Table 3. The clumps were then identified and masked out to produce a clump-free image. See Fig. 4 for the projected temperature, deprojected density, and 3D entropy profiles constructed from the resulting surface brightness profiles and spectral fits. The results for the 3.0 keV clumps were then used in the entropy-correction process for 0.7 keV clumps and to determine number density values, as outlined in Section 2.3.

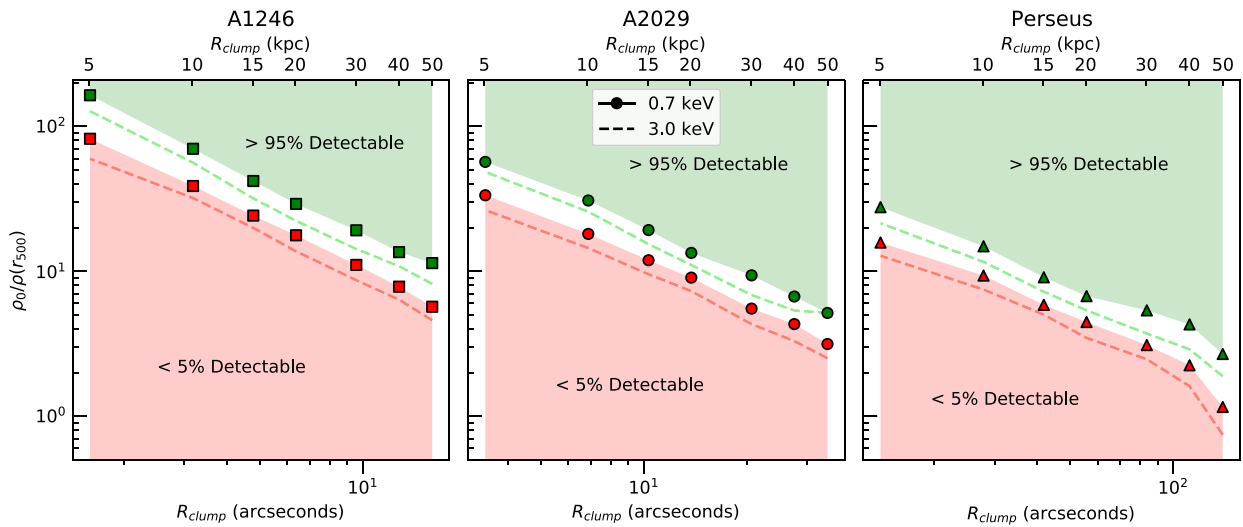

**Figure 3.** Detectability ranges of gas clumps with central temperatures of 0.7 keV (points) and 3.0 keV (dashed lines) based on STAR-X's instrumental background and the galactic foreground. The green points and the dashed lines represent central clump densities that are approximately 95 per cent detectable, while the red points and lines represent central clump densities that are approximately 5 per cent detectable. The central density of the clumps are scaled by the density at $r_{500}$ for each cluster. The green regions highlight the parameter space where clumps with a central temperature of 0.7 keV are extremely likely to be detected by STAR-X, greater than 95 per cent detectable. The red regions highlight the parameter space where 0.7 keV clumps are extremely unlikely to be detected, less than 5 per cent detectable.

The entropy profile was calculated using the 3D temperature model used to construct the cluster, instead of the extracted, projected temperatures. Iterative fitting of the projected surface brightness and temperature profiles, as done for actual observations, is more involved and not crucial for our aims as long as the measured projected temperatures match the projected temperatures produced by the 3D model, which is the case.

### 4.2 Abell 1246

The same recreation and analysis process in Section 2 was carried out for Abell 1246, also for a 100 ks mock *STAR-X* observation. Temperature and emissivity profiles were obtained by manually fitting equations (2) and (3) to observational data obtained from Sato et al. (2014). See Tables 1 and 2 for the density and temperature parameters used to initially recreate the cluster, which were verified to match observational data. The temperature and density profiles were then modified in the outskirts to yield the expected entropy profile; the resulting parameters can be found in Tables 1 and 2. The entropy-correction process was carried out for two central clump temperature assessments; 3.0 and 0.7 keV. For a detailed description of the correction process, see Section 2.3. Clumps with a central temperature of 3.0 keV were injected into their respective, corrected recreation to obtain number density values for a range of clump sizes and their respective central densities. The results for the 3.0 keV central clump temperature analysis were then used in the entropy-correction process to assess 0.7 keV clumps and to determine their matching number density values, as outlined in Section 2.3. The projected temperatures, deprojected densities, and derived entropy profiles can be found in Fig. 5. The identified number density values for the clumps can be found in Table 3. The methods for obtaining number density values for clumps and deriving the entropy profile are identical to the methods used in the previous subsection for Abell 2029.

### 4.3 Perseus cluster

With a smaller redshift, the Perseus cluster would require multiple pointings. As an approximation of an observation from *STAR-X*, the FOV in the simulation was increased to cover the entirety of the Perseus Cluster. Thus, a true mosaic was not produced as would be in a real observation, although the outcome is equivalent since off-axis variations in PSF and effective area are not included in the simulations. The larger FOV mock observation was equivalent to ∼9 pointings, each with a *STAR-X* exposure time of 100 ks. The Perseus cluster was recreated using observed profiles from Simionescu et al. (2011) and Urban et al. (2013). See Tables 1 and 2 for the density and temperature parameters used to initially recreate the cluster, which were verified to match observational data. The temperature and density profiles were adjusted in the outskirts to yield the expected entropy profile. This correction process was used to assess two gas clump temperatures; 3.0 and 0.7 keV. See Section 2.3 for details. The corrected profile parameters can be found in Tables 1 and 2. Gas clumps with properties identified in Table 3 for a central temperature of 3.0 keV were then injected into the cluster recreation to obtain number density values that recreate the observed drop in entropy for the Perseus Cluster. The results from the 3.0 keV analysis were then used in the entropy-correction process intended for gas clumps with a central temperature of 0.7 keV and to determine number density values, as outlined in Section 2.3. Projected temperatures, deprojected densities, and derived entropy profiles are shown in Fig. 6. The identified number density values can be found in Table 3. The methods for obtaining number density values for clumps and deriving the entropy profile are identical to the methods used for Abell 2029.

## 5 DISCUSSION

The main assumption in this work has been that clumping occurs in the outskirts of galaxy clusters, which then causes observed entropy profiles to unexpectedly drop off in the outskirts. This assumption is supported by simulations (Walker et al. 2019) and the idea that a galaxy cluster would not stay in a state with a sustained lower temperature and higher density in the outskirts if it is in hydrostatic equilibrium (Biffi et al. 2016). Thus, a phenomena like gas clumping is potentially necessary to explain the unexpected drop in entropy profiles that are consistently observed in the outskirts of galaxy clusters. See Section 1 for a more in-depth description and example of clusters with observed drops in entropy.

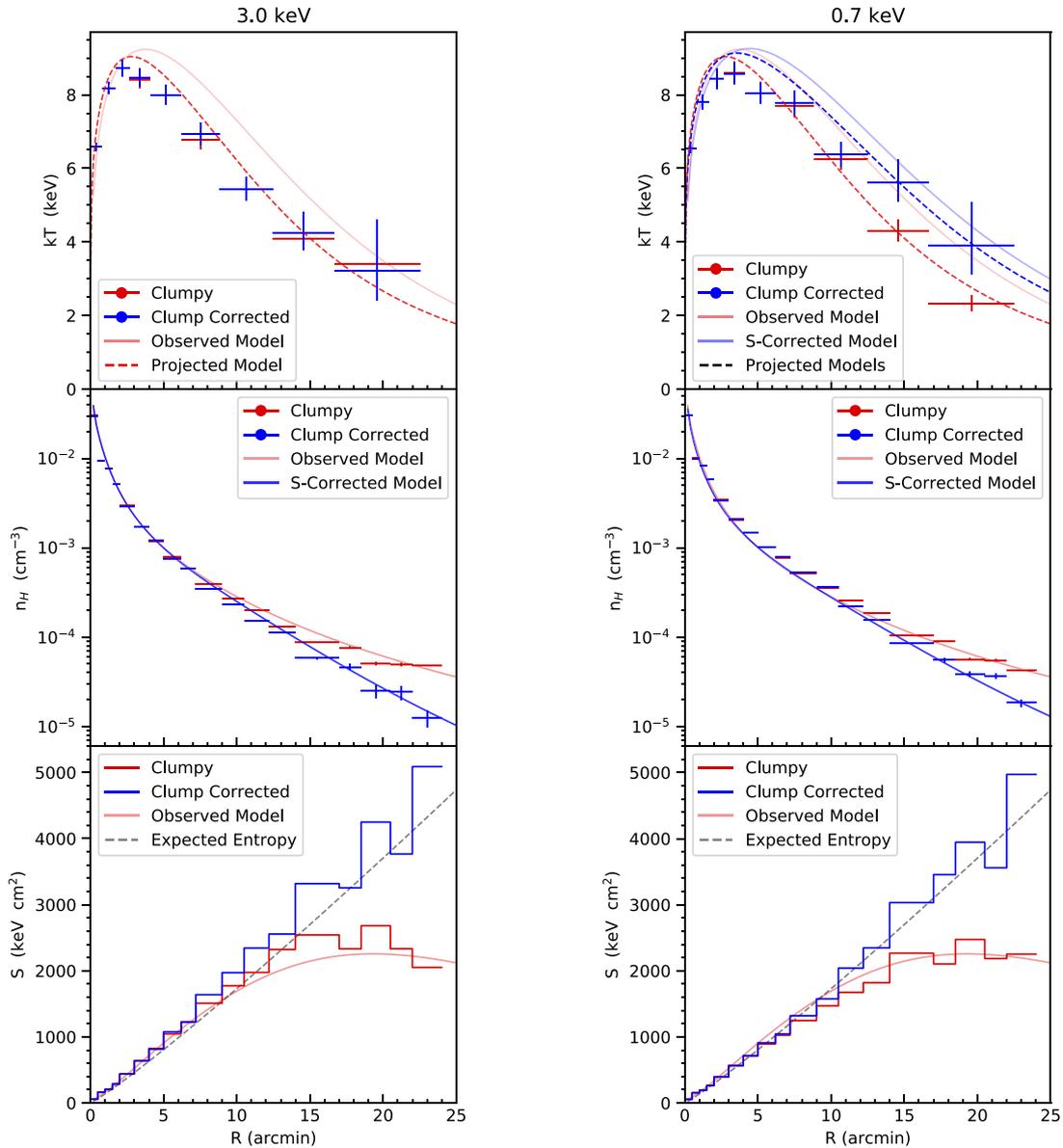

**Figure 4.** Top: Projected temperatures for an Abell 2029 recreation with injected gas clumps with a central temperature of 3.0 keV (left) and 0.7 keV (right). Results of the injected clumps can be seen in red, while clump-corrected results can be seen in blue. The dashed lines signify projected temperature profiles, while the solid lines represent best-fitting 3D models. Right panel includes an entropy-corrected temperature profile (blue), which was used to construct the cluster. Middle: Deprojected hydrogen number density profiles for the clumpy A2029 recreation (red) and the clump-corrected results (blue). The blue line represents the density profile necessary to yield the expected entropy and was used to construct the cluster. In the right panel, the entropy-corrected density (blue) is tied to the entropy-corrected temperature profile above (blue) to yield the expected entropy. Bottom: Derived entropy profiles for the clumpy A2029 recreation and accompanying clump-correction (blue). The dashed grey line signifies the expected entropy ($r^{1.1}$) relation. The model fit to observations (pink) is identical in left and right panels for each quantity measured.

The goal of this work has been to assess how *STAR-X* would behave if gas clumping occurs in the outskirts of galaxy clusters. It is proposed that previous missions either lacked the sensitivity or spatial resolution to directly detect the clumps, which resulted in a bias in the derived entropy profiles.

This work attempted to assess the properties of gas clumps that would be necessary to recreate observed drops in entropy. The sizes of the clumps used across all of the simulations ($R_{\rm clump}$ = 5–50 kpc) were chosen to match clump sizes found in cosmological simulations, taken from Vazza et al. (2012) along with the central clump temperature of 3.0 keV, which had a high differential distribution in simulations. A central clump temperature of 0.7 keV was also assessed. First, clumps that were bright enough to be detectable by

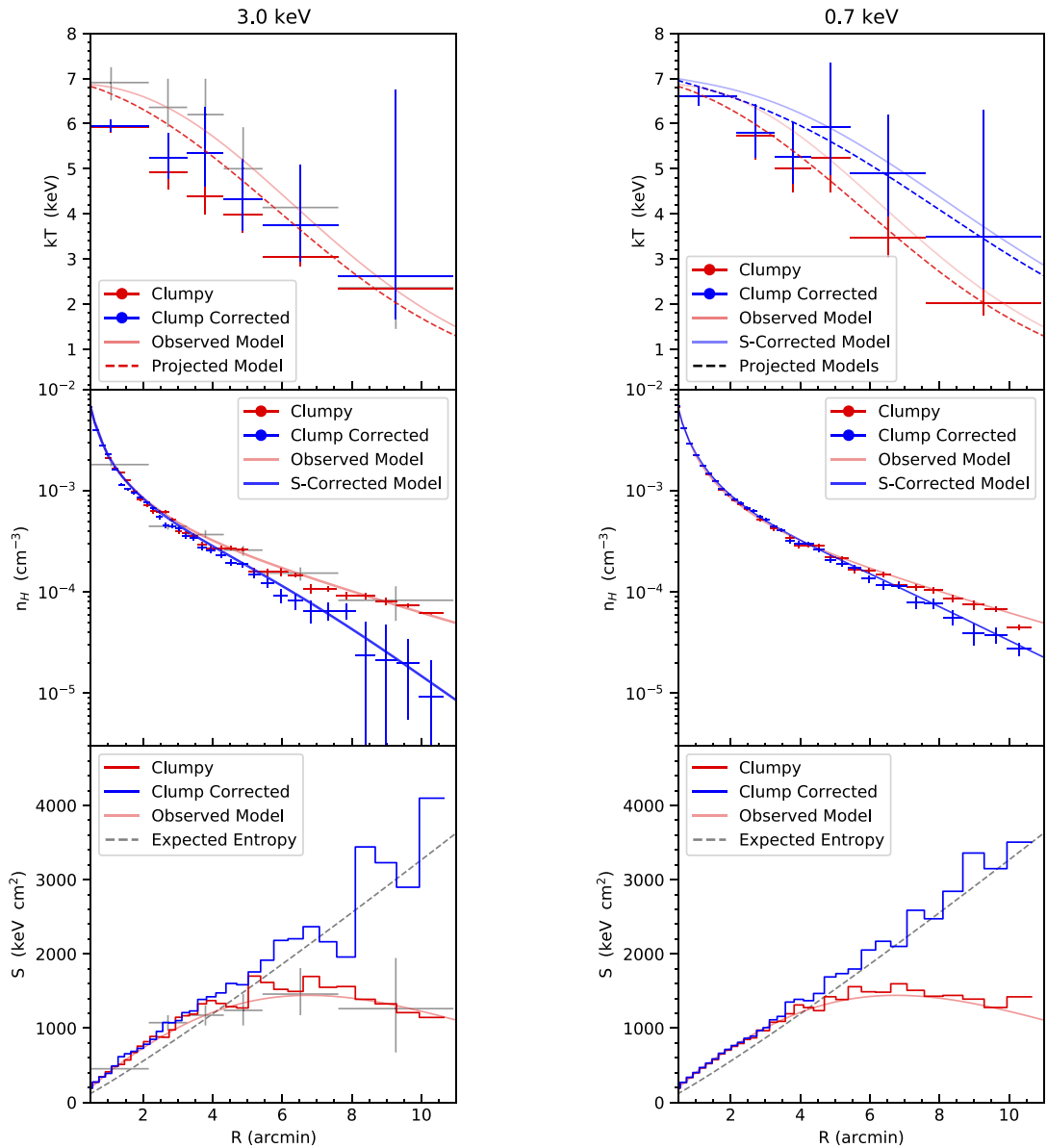

**Figure 5.** Top: Projected temperatures for an Abell 1246 recreation with injected gas clumps with a central temperature of 3.0 keV (left) and 0.7 keV (right). Results of the injected clumps can be seen in red, while clump-corrected results can be seen in blue. The dashed lines signify projected temperature profiles, while the solid lines represent best-fitting 3D models. Right panel includes an entropy-corrected temperature profile (blue), which was used to construct the cluster. Middle: Deprojected hydrogen number density profiles for the clumpy A1246 recreation (red) and the clump-corrected results (blue). The blue line represents the density profile necessary to yield the expected entropy and was used to construct the cluster. In the right panel, the entropy-corrected density (blue) is tied to the entropy-corrected temperature profile above (blue) to yield the expected entropy. Bottom: Derived entropy profiles for the clumpy A1246 recreation and accompanying clump-correction (blue). The dashed grey line signifies the normalized, expected entropy ($r^{1.1}$) relation. The model fit to observations (pink) is identical in left and right panels for each quantity measured. The grey points in left panels are observational data used to model the cluster recreation (Sato et al. 2014).

*STAR-X* were identified for a range of clump sizes by establishing a detectability threshold based on clump density and size at different cluster redshifts. This was done to explore a range of clump properties that *STAR-X* would be able to detect, which can be seen in Fig. 3. The central density $\rho_0$ of the clumps, relative to the density of the ambient ICM at $r_{500}$, is compared to the size of the clumps $R_{clump}$ in terms of *STAR-X*'s ability to successfully detect and mask out the clumps. In the green region of Fig. 3, clumps with these properties will be easily identifiable with *STAR-X*, with >95 per cent of clump emission being masked out; the points at the edge of this region indicate the clump properties used in our simulations. The red region indicates the range of properties that clumps would need to have in order to continue to avoid detection.

It is worth noting that in actual observations, the sizes of the clumps would be unknown. In the detectability analysis, the known clump size is used to calculate a projected scale size, which is fed to `wavdetect`. It is in this fashion that only one clump size is searched for in each observation. In the detectability analysis, we found that a small portion of a larger clump could be falsely identified when searching at a smaller scale, leaving the rest of the clump undetected. It is for this reason that the detectability analysis in Section 3.1 uses a single scale size and the point source detection and clump detection are done separately in Section 4, to allow `wavdetect` to search for only one scale size at a time. It may be feasible to search an actual observation for different clump sizes individually to account for this discrepancy, rather than feeding `wavdetect` a range of scale sizes all at once.

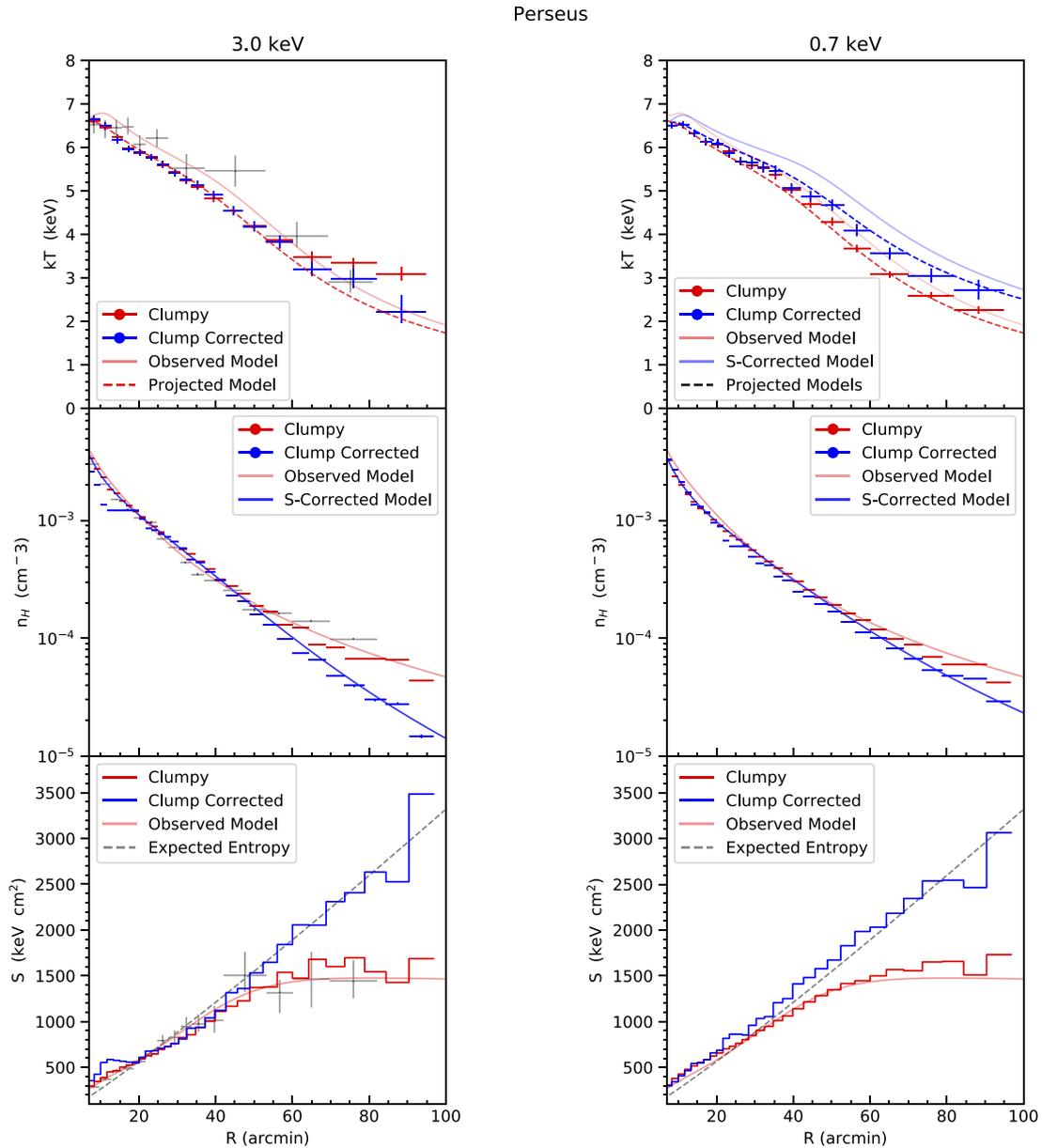

**Figure 6.** Top: Projected temperatures for a Perseus Cluster recreation with injected gas clumps with a central temperature of 3.0 keV (left) and 0.7 keV (right). Results of the injected clumps can be seen in red, while clump-corrected results can be seen in blue. The dashed lines signify projected temperature profiles, while the solid lines represent best-fitting 3D models. Right panel includes an entropy-corrected temperature profile (blue), which was used to construct the cluster. Middle: Deprojected hydrogen number density profiles for the clumpy Perseus recreation (red) and the clump-corrected results (blue). The blue line represents the density profile necessary to yield the expected entropy and was used to construct the cluster. In the right panel, the entropy-corrected density (blue) is tied to the entropy-corrected temperature profile above (blue) to yield the expected entropy. Bottom: Derived entropy profiles for the clumpy Perseus recreation and accompanying clump-correction (blue). The dashed grey line signifies the normalized, expected entropy ($r^{1.1}$) relation. The model fit to observations (pink) is identical in left and right panels for each quantity measured. The grey points in left panels are observational data used to model the cluster recreation (Urban et al. 2013).

With an established range of detectable clump properties, observed drops in entropy were recreated by injecting clumps into entropy-corrected versions of observed clusters. Once sufficient drops were observed through mock observations, *STAR-X*'s ability to resolve gas clumps was assessed.

For clumps with a central temperature of 0.7 keV, a slightly higher central density was necessary to produce clumps that were as detectable compared to clumps with a central temperature of 3.0 keV, as determined in Section 3 and seen in Fig. 3.

A higher central density implied that a lower number density of clumps to bias entropy profiles is necessary, as seen in Table 3.

The entropy-correction for clumps with a central temperature of 0.7 keV yielded cluster temperature profiles that were hotter in the outskirts with corresponding density profiles that were less dense in the outskirts. Combining these profiles to calculate entropy (equation 1) resulted in an approximation of the expected entropy profile. The resulting density profiles were slightly denser than the entropy-corrected profiles intended for clumps with a central temperature of 3.0 keV clumps. This makes sense because for the 3.0 keV clumps, the cluster temperature profiles were not modified. Therefore, for the lower clump temperature of 0.7 keV, which included a warmer temperature cluster profile, the density profile would have to be slightly higher to achieve the same entropy profile achieved with the 3.0 keV clumps.

Three relaxed clusters with observed drops in entropy in their outskirts were chosen for reconstruction: Abell 2029, Abell 1246, and the Perseus Cluster. Each cluster is in a relaxed dynamical state. Abell 2029 and Abell 1246 were chosen because they fit entirely within *STAR-X*'s FOV and could be observed in one pointing. This is an efficient characteristic of *STAR-X* as it has a wide FOV. The Perseus cluster is an alternative case. Due to its low redshift, multiple pointings would be required for a complete observation. Because it is in closer proximity, fainter clumps could theoretically be detectable in the Perseus cluster.

Abell 2029 was observed by the *Suzaku* observatory to have a significant drop in entropy in the outskirts (Walker et al. 2012a). With an ample number of clumps randomly distributed within an approximation of an entropy-corrected Abell 2029, a similar drop in entropy can be recreated. This is seen in Fig. 4. Clumping scenarios that are detectable by *STAR-X* and recreate an observed drop in entropy can be found in Table 3. The number density of clumps within the cluster is identified through the reconstructions and analysis. It is apparent that for Abell 2029, if gas clumping is a viable explanation for the apparent drop in entropy and clumps are within the bounds of detectability established in Section 3.1, *STAR-X* would be sufficiently sensitive enough for the clumps to be identified and subtracted out in order to recover the expected entropy profile (the relation identified in Voit et al. 2005).

Abell 1246 was chosen as a cluster to reconstruct as it is another plausible candidate for *STAR-X* that fully fits within the FOV and has had an observed drop in entropy in the outskirts (Sato et al. 2014). It has a slightly lower temperature than Abell 2029. The results for Abell 1246 follow those of Abell 2029. Once enough clumps are injected into the cluster, a drop in entropy equivalent to that seen in observations can be recreated. With sufficiently dense clumps, *STAR-X* is able to identify and mask out clumps to recover the expected entropy profile. See Fig. 5 for the relevant temperature, density, and entropy profiles and Table 3 for the established clump properties.

The Perseus cluster was selected to provide an example of a cluster with a lower redshift that would require multiple observations to fully encompass it. It was also chosen to examine scenarios in which fainter clumps could be detected due to the lower redshift of the cluster. Clumps that are 95 per cent detectable by *STAR-X* for the Perseus Cluster therefore had much lower density values for a range of clump sizes, as demonstrated in Table 3. For clump sizes that are less than 20 kpc, the number density of detectable clumps that recreate an observed drop in entropy is so great that during the correction process, much of the cluster's counts in the outskirts are removed. It was therefore difficult to fully recover the expected entropy profile for larger number densities as the number of counts per annulus was so low. For clumps larger than 20 kpc, with a density on the detectability threshold for *STAR-X* at the Perseus Cluster's redshift, clumps could be masked out without removing so much solid angle that the emission from the diffuse ICM could not be constrained. Every clump size was able to recreate the observed drop in this cluster. However, the sheer number of clumps required at a lower density and radial size yielded potentially unrealistic physical situations.

Each of the three clusters chosen for recreation produced similar results. With established brightness thresholds for gas clumps at each of the cluster's redshifts, a number density of clumps that recreated an observed drop in entropy could be identified. In each case, *STAR-X* was able to identify clumps sufficiently enough to recover the expected entropy profile.

For each cluster, there was an inner radius within which no clump photons were detected as the cluster emission is much brighter in the centre. In cluster outskirts where the emission was much fainter, a radius was also present within each cluster where beyond it clumps were 90($\pm$5) per cent detectable. Beyond this radius clumps were considered to be fully detectable. For clusters corrected for gas clumps with a central temperature of 0.7 keV, clump photons were 0 per cent detectable within 7.2, 2.7, and 10 arcmin for Abell 2029, Abell 1246, respectively, and 90($\pm$5) per cent detectable beyond 16.7, 6.5, and 63 arcmin for Abell 2029, Abell 1246, and the Perseus Cluster, respectively. For clusters corrected for gas clumps with a central temperature of 3.0 keV, clump photons were 0 per cent detectable within 5.2, 2.5, and 9.1 arcmin for Abell 2029, Abell 1246, and the Perseus Cluster, respectively and 90($\pm$5) per cent detectable beyond 14.7, 6.4, and 59 arcmin for Abell 2029, Abell 1246, and the Perseus Cluster, respectively.

For clumps that were at the lower end of the detectability range (around 5 per cent detectable), clump-subtraction and entropy correction have little effect. Since they cannot be masked out, the clumps still impact the entropy profile heavily. At 5 per cent detectability, the number of clumps required to recreate an observed drop in entropy is roughly four times that of the 95 per cent detectable clumps.

An overabundance of clumps is not necessarily realistic. At a certain point, if the ICM is saturated with pockets of denser and cooler structure, the 'clumps' become the ICM. In any case, the large number of clumps would create small-scale surface brightness fluctuations that could be detected statistically. Zhuravleva et al. (2015) relates gas clumps to large scale density fluctuations in the ICM, which could imply larger clumps being more probable. It is also possible that gas clumps do not exist in the outskirts of clusters and the drop in entropy that is observed is caused by something else entirely, such as projected large-scale structures along the line of sight, such as more distant background clusters or foreground groups. These objects can be thought of as large clumps, and given the trend in Fig. 3, they should be detectable even with apparent central densities comparable to the ICM outskirts.

Alternatively, Walker et al. (2012b) proposes that a drop in entropy is caused by a reduction in accretion shock strength, in agreement with Cavaliere et al. (2011) and Lapi, A. et al. (2010). A lack of observed clumps in real *STAR-X* data would be in support of this scenario.

Since *STAR-X*'s FOV and other capabilities are similar to that of *Athena*'s Wide Field Imager (Barcons 2015), these results can generally be applied to *Athena* observations of nearby clusters as well.

## 6 CONCLUSION

The primary goal of this research was to determine whether or not *STAR-X*, an observatory proposed to the MIDEX call by NASA in 2021, would be sensitive enough to detect and mask out gas clumps in the outskirts of galaxy clusters in order to recover entropy profiles of the truly diffuse ICM. We first determined what properties gas clumps needed to have in order to be detectable by *STAR-X*. A detectability threshold based on their central gas density was identified for a range of clump sizes at the specific cluster redshifts considered here. This allowed for an establishment of properties of the dimmest gas clumps that would be detectable by *STAR-X* in these clusters. Then the goal was to identify how many of each detectable clump were necessary to recreate a drop in entropy in the outskirts of our reconstructed galaxy clusters equivalent to that observed in the real clusters. Mock observations included the Galactic foreground and instrumental background of *STAR-X*, as well as randomly generated point sources following the CXB. These mock observations were then analysed using `wavdetect` to subtract out point sources and gas clumps and two radial entropy profiles were extracted to assess the impact of the gas clumps and the success of clump identification for the derived entropy profile in the mock observations. Abell 2029, Abell 1246, and the Perseus Cluster were chosen for reconstruction and in each case, clump properties that recreated observed drops in entropy could be identified. For the majority of identified clump properties, *STAR-X* was able to identify and mask out gas clumps in order to recover the expected entropy profiles. A parameter space for clumps that are both detectable by *STAR-X* and which can recreate observed drops in entropy were subsequently identified through this process. If gas clumps are as prominent as they have been found to be in simulations (Vazza et al. 2012) and in this work, *STAR-X* would allow for a much deeper understanding of cluster outskirts.

## ACKNOWLEDGEMENTS

We thank the anonymous referees for their constructive feedback that helped to improve the paper.

## DATA AVAILABILITY

The data underlying this article will be shared on reasonable request to the corresponding author.

## SUPPORTING INFORMATION

Supplementary data are available at *RASTAI* online.

**STAR-X Clumping Paper for RASTI (Final).zip**

Please note: Oxford University Press is not responsible for the content or functionality of any supporting materials supplied by the authors. Any queries (other than missing material) should be directed to the corresponding author for the article.

This paper has been typeset from a T<sub>E</sub>X/L<sup>A</sup>T<sub>E</sub>X file prepared by the author.